\documentclass[10pt,a4paper,onecolumn]{article}
\usepackage{marginnote}
\usepackage{graphicx}
\usepackage{xcolor}
\usepackage{authblk,etoolbox}
\usepackage{titlesec}
\usepackage{calc}
\usepackage{tikz}
\usepackage{hyperref}
\hypersetup{colorlinks,breaklinks=true,
            urlcolor=[rgb]{0.0, 0.5, 1.0},
            linkcolor=[rgb]{0.0, 0.5, 1.0},
            citecolor = blue}
\usepackage{caption}
\usepackage{tcolorbox}
\usepackage{amssymb,amsmath}
\usepackage{ifxetex,ifluatex}
\usepackage{seqsplit}
\usepackage{xstring}
\usepackage{natbib}
\usepackage[T1]{fontenc}
\usepackage[utf8]{inputenc}
\usepackage{cleveref}
\usepackage{float}
\usepackage{orcidlink}
\usepackage{xspace}
\usepackage{listings}
\usepackage{comment}
\usepackage{longtable,booktabs,array}
\usepackage{tikz}
\usetikzlibrary{shapes,arrows,shadows,fit}
\usetikzlibrary{positioning}
\usetikzlibrary{bayesnet}

\let\origfigure\figure
\let\endorigfigure\endfigure
\renewenvironment{figure}[1][2] {
    \expandafter\origfigure\expandafter[H]
} {
    \endorigfigure
}

\let\textttOrig=\texttt
\def\texttt#1{\expandafter\textttOrig{\seqsplit{#1}}}
\renewcommand{\seqinsert}{\ifmmode
  \allowbreak
  \else\penalty6000\hspace{0pt plus 0.02em}\fi}

\usepackage{abstract}

\makeatletter
\let\href@Orig=\href
\def\href@Urllike#1#2{\href@Orig{#1}{\begingroup
    \def\Url@String{#2}\Url@FormatString
    \endgroup}}
\def\href@Notdoi#1#2{\def\tempa{#1}\def\tempb{#2}%
  \ifx\tempa\tempb\relax\href@Urllike{#1}{#2}\else
  \href@Orig{#1}{#2}\fi}
\def\href#1#2{%
  \IfBeginWith{#1}{https://doi.org}%
  {\href@Urllike{#1}{#2}}{\href@Notdoi{#1}{#2}}}
\makeatother

\newlength{\cslhangindent}
\newlength{\csllabelwidth}
 {
  \setlength{\parindent}{0pt}
  \ifodd #1 \everypar{\setlength{\hangindent}{\cslhangindent}}\ignorespaces\fi
  \ifnum #2 > 0
  \setlength{\parskip}{#2\baselineskip}
  \fi
 }%
 {}
\usepackage{calc}

\usepackage[top=3.5cm, bottom=3cm, right=1.5cm, left=1.0cm,
            headheight=2.2cm, reversemp, includemp, marginparwidth=4.5cm]{geometry}

\titleformat{\section}
  {\normalfont\sffamily\Large\bfseries}
  {}{0pt}{}
\titleformat{\subsection}
  {\normalfont\sffamily\large\bfseries}
  {}{0pt}{}
\titleformat{\subsubsection}
  {\normalfont\sffamily\bfseries}
  {}{0pt}{}
\titleformat*{\paragraph}
  {\sffamily\normalsize}

\usepackage{fancyhdr}
\makeatletter
\let\ps@plain\ps@fancy
\fancyheadoffset[L]{4.5cm}
\fancyfootoffset[L]{4.5cm}

\definecolor{linky}{rgb}{0.0, 0.5, 1.0}

\newtcolorbox{repobox}
   {colback=red, colframe=red!75!black,
     boxrule=0.5pt, arc=2pt, left=6pt, right=6pt, top=3pt, bottom=3pt}

\newcommand{\ExternalLink}{%
   \tikz[x=1.2ex, y=1.2ex, baseline=-0.05ex]{%
       \begin{scope}[x=1ex, y=1ex]
           \clip (-0.1,-0.1)
               --++ (-0, 1.2)
               --++ (0.6, 0)
               --++ (0, -0.6)
               --++ (0.6, 0)
               --++ (0, -1);
           \path[draw,
               line width = 0.5,
               rounded corners=0.5]
               (0,0) rectangle (1,1);
       \end{scope}
       \path[draw, line width = 0.5] (0.5, 0.5)
           -- (1, 1);
       \path[draw, line width = 0.5] (0.6, 1)
           -- (1, 1) -- (1, 0.6);
       }
   }

\patchcmd{\@maketitle}{center}{flushleft}{}{}
\patchcmd{\@maketitle}{center}{flushleft}{}{}
\patchcmd{\@maketitle}{\LARGE}{\LARGE\sffamily}{}{}

\def\maketitle{{%

  \ifdefined\collaborationline
  \g@addto@macro\AB@authors{%
    \protect\\[0.5em]%
    \protect\normalfont\protect\itshape\collaborationline%
  }%
  \fi
  \AB@maketitle}}

\renewcommand\AB@affilsepx{ \protect\Affilfont}
\renewcommand\AB@affilnote[1]{{\bfseries #1}\hspace{3pt}}
\renewcommand{\affil}[2][]%
   {\newaffiltrue\let\AB@blk@and\AB@pand
      \if\relax#1\relax\def\AB@note{\AB@thenote}\else\def\AB@note{#1}%
        \setcounter{Maxaffil}{0}\fi
        \begingroup
        \let\href=\href@Orig
        \let\texttt=\textttOrig
        \let\protect\@unexpandable@protect
        \def\thanks{\protect\thanks}\def\footnote{\protect\footnote}%
        \@temptokena=\expandafter{\AB@authors}%
        {\def\\{\protect\\\protect\Affilfont}\xdef\AB@temp{#2}}%
         \xdef\AB@authors{\the\@temptokena\AB@las\AB@au@str
         \protect\\[\affilsep]\protect\Affilfont\AB@temp}%
         \gdef\AB@las{}\gdef\AB@au@str{}%
        {\def\\{, \ignorespaces}\xdef\AB@temp{#2}}%
        \@temptokena=\expandafter{\AB@affillist}%
        \xdef\AB@affillist{\the\@temptokena \AB@affilsep
          \AB@affilnote{\AB@note}\protect\Affilfont\AB@temp}%
      \endgroup
       \let\AB@affilsep\AB@affilsepx
}
\makeatother

\renewcommand\Affilfont{\sffamily\small\mdseries}
\ifnum 0\ifxetex 1\fi\ifluatex 1\fi=0 
  \usepackage[T1]{fontenc}
  \usepackage[utf8]{inputenc}

\else 
  \ifxetex
    \usepackage{mathspec}
    \usepackage{fontspec}

  \else
    \usepackage{fontspec}
  \fi
  \defaultfontfeatures{Ligatures=TeX,Scale=MatchLowercase}

\fi
\IfFileExists{upquote.sty}{\usepackage{upquote}}{}
\IfFileExists{microtype.sty}{%
\usepackage{microtype}
\UseMicrotypeSet[protrusion]{basicmath} 
}{}

\let\addcontentslineOrig=\addcontentsline
\def\addcontentsline#1#2#3{\bgroup
  \let\texttt=\textttOrig\addcontentslineOrig{#1}{#2}{#3}\egroup}
\let\markbothOrig\markboth
\def\markboth#1#2{\bgroup
  \let\texttt=\textttOrig\markbothOrig{#1}{#2}\egroup}
\let\markrightOrig\markright
\def\markright#1{\bgroup
  \let\texttt=\textttOrig\markrightOrig{#1}\egroup}

\usepackage{graphicx,grffile}
\makeatletter
\def\maxwidth{\ifdim\Gin@nat@width>\linewidth\linewidth\else\Gin@nat@width\fi}
\def\maxheight{\ifdim\Gin@nat@height>\textheight\textheight\else\Gin@nat@height\fi}
\makeatother
\setkeys{Gin}{width=\maxwidth,height=\maxheight,keepaspectratio}
\IfFileExists{parskip.sty}{%
\usepackage{parskip}
}{
\setlength{\parindent}{0pt}
\setlength{\parskip}{6pt plus 2pt minus 1pt}
}
\providecommand{\tightlist}{%
  \setlength{\itemsep}{0pt}\setlength{\parskip}{0pt}}

\ifx\paragraph\undefined\else
\let\oldparagraph\paragraph
\renewcommand{\paragraph}[1]{\oldparagraph{#1}\mbox{}}
\fi
\ifx\subparagraph\undefined\else
\let\oldsubparagraph\subparagraph
\renewcommand{\subparagraph}[1]{\oldsubparagraph{#1}\mbox{}}
\fi

\DeclareFixedFont{\ttb}{T1}{txtt}{bx}{n}{10} 
\DeclareFixedFont{\ttm}{T1}{txtt}{m}{n}{10}  

\usepackage{color}
\definecolor{deepblue}{rgb}{0,0,0.5}
\definecolor{deepred}{rgb}{0.6,0,0}
\definecolor{deepgreen}{rgb}{0,0.5,0}

\usepackage{listings}

\newcommand\pythonstyle{\lstset{
language=Python,
basicstyle=\scriptsize,
morekeywords={self},              
keywordstyle=\scriptsize\color{deepblue},
emph={MyClass,__init__},          
emphstyle=\ttb\color{deepred},    
stringstyle=\color{deepgreen},
    numbers=left,
frame=tb,                         
showstringspaces=false
}}

\lstnewenvironment{python}[1][]
{
\pythonstyle
\lstset{#1}
}
{}

\newcommand\pythoninline[1]{{\pythonstyle\lstinline!#1!}}

\newcommand{\pkgfont}{\texttt} 
\newcommand{\pkg}[2][]{%
  \if\relax\detokenize{#1}\relax
    \pkgfont{#2}%
  \else
    \href{#1}{\pkgfont{#2}}%
  \fi
}

\usepackage{listings}
\usepackage{xcolor}

\lstdefinelanguage{YAML}{
  keywords={true, false, null},
  keywordstyle=\color{blue}\bfseries,
  comment=[l]{\#},
  commentstyle=\color{gray}\itshape,
  stringstyle=\color{teal},
  morestring=[b]",
  morestring=[b]',
}

\title{\texttt{cosmicjellyfish}: Fisher-matrix and Monte Carlo forecasts
for cosmological surveys in Python and JAX}

\date{\vspace{-7ex}}

\begin{document}
\author[1, 2]{Santiago
Casas\thanks{Corresponding author}\,\orcidlink{0000-0002-4751-5138}}
\author[3]{Sefa Pamuk\,\orcidlink{0009-0004-0852-8624}}
\author[2, 4]{V. M. Sabarish\,\orcidlink{0000-0001-5677-0838}}
\author[18, 19, 20]{Ziad Sakr\,\orcidlink{0000-0002-4823-3757}}
\author[5]{Domenico Sapone\,\orcidlink{0000-0001-7089-4503}}
\author[6, 7]{Matías Leizerovich\,\orcidlink{0000-0002-6438-2285}}
\author[8, 9]{Noemi Frusciante\,\orcidlink{0000-0002-7375-1230}}
\author[10, 11, 12, 13]{Francesco Pace\,\orcidlink{0000-0001-8039-0480}}
\author[15]{Emilio Bellini\,\orcidlink{0000-0003-4762-0795}}
\author[14]{Kazuya Koyama\,\orcidlink{0000-0001-6727-6915}}
\author[10, 11, 12, 16]{Stefano Camera\,\orcidlink{0000-0003-3399-3574}}
\author[17]{Marco Raveri\,\orcidlink{0000-0002-7354-3802}}

\affil[1]{Institute for Theoretical Particle Physics and Cosmology
(TTK), RWTH Aachen University, Germany}
\affil[2]{Scientific Information, German Aerospace Center (DLR),
Germany}
\affil[3]{Instituto de Física de Cantabria (IFCA), Spain}
\affil[4]{Hamburger Sternwarte, Universität Hamburg, Germany}
\affil[5]{Departamento de Física, FCFM, Universidad de Chile, Chile}
\affil[6]{Departamento de Física, Facultad de Ciencias Exactas y
Naturales, Universidad de Buenos Aires, Argentina}
\affil[7]{Instituto de Física de Buenos Aires (IFIBA), CONICET --
Universidad de Buenos Aires, Argentina}
\affil[8]{Dipartimento di Fisica ``E. Pancini'', Università degli Studi
di Napoli ``Federico II'', Italy}
\affil[9]{INFN -- Sezione di Napoli, Italy}
\affil[10]{Dipartimento di Fisica, Università degli Studi di Torino,
Italy}
\affil[11]{INFN -- Sezione di Torino, Italy}
\affil[12]{INAF -- Osservatorio Astrofisico di Torino, Italy}
\affil[13]{Instituto de Astrofísica e Ciências do Espaço, Faculdade de
Ciências, Universidade de Lisboa, Portugal}
\affil[14]{Institute of Cosmology and Gravitation, University of
Portsmouth, United Kingdom}
\affil[15]{Center for Astrophysics and Cosmology, University of Nova
Gorica, Slovenia}
\affil[16]{Department of Physics and Astronomy, University of the
Western Cape, South Africa}
\affil[17]{Department of Physics, University of Genova, and INFN --
Sezione di Genova, Italy}
\affil[18]{Instituto de Física Teórica, Spain}
\affil[19]{Institut de Recherche en Astrophysique et Planétologie,
France}
\affil[20]{Université Saint Joseph, Lebanon}

\maketitle

%
\newcommand{\sidebaroffset}{-260pt}
\marginpar{
  \begin{flushleft}
  \vspace{\sidebaroffset}
  \sffamily\small

  {\bfseries DOI: N/A}

  \vspace{2mm}

  {\bfseries Software}
  \begin{itemize}
    \setlength\itemsep{0em}
    \item \href{https://github.com/santiagocasas/cosmicjellyfish}{\color{linky}{Repository}} \ExternalLink
    \item \href{https://www.santicasas.space/cosmicjellyfish/docs/}{\color{linky}{Documentation}} \ExternalLink
    \item \href{https://www.santicasas.space/cosmicjellyfish/dashboard/}{\color{linky}{Validation dashboard}} \ExternalLink
    \item \href{https://huggingface.co/datasets/santicas/cjfp-data}{\color{linky}{Validation data}} \ExternalLink
    \item \href{https://doi.org/10.5281/zenodo.24593176}{\color{linky}{Archive (Zenodo)}} \ExternalLink
  \end{itemize}

  \vspace{2mm}
  {\bfseries Install:}\\ \textttOrig{pip install cosmicjellyfish}

  \vspace{2mm}
  \par\noindent\hrulefill\par
  \vspace{2mm}

  {\bfseries Submitted:} N/A \\
  {\bfseries Published:} N/A

  \vspace{2mm}
  {\bfseries License}\\
  Authors retain copyright and release the work under CC BY 4.0
  (\href{http://creativecommons.org/licenses/by/4.0/}{\color{linky}{link}}).
  The software is released under the Apache 2.0 licence.
  \end{flushleft}
}

\vspace{.2cm}

\section{Summary}\label{summary}

\texttt{cosmicjellyfish} is a Python package for Fisher-matrix and Monte
Carlo forecasts of cosmological surveys. It is the current version of
\texttt{cosmicfishpie}, the Python successor of the
\href{https://github.com/CosmicFish/CosmicFish}{CosmicFish} code
\citep{Raveri2016}, and has been used in about ten publications since
the Euclid forecast-validation programme \citep{Blanchard2020}. It
computes the observables of photometric galaxy clustering and weak
lensing, spectroscopic galaxy clustering, 21cm intensity mapping and the
cosmic microwave background, and returns the expected parameter
constraints either as a Fisher matrix or as a posterior obtained by
nested sampling with
\href{https://nautilus-sampler.readthedocs.io/en/latest/}{Nautilus},
both from a single configuration, so that going beyond the Gaussian
approximation for a forecast, is only a matter of computing time. Theory
predictions are supplied interchangeably by
\href{https://github.com/cmbant/CAMB}{CAMB},
\href{https://github.com/lesgourg/class_public}{CLASS},
\href{https://github.com/DeaglanBartlett/symbolic_pofk}{symbolic
emulators} \citep{Bartlett2024a, Bartlett2024b, Sui2025}, or
\href{https://github.com/DifferentiableUniverseInitiative/jax_cosmo}{jax-cosmo};
derivatives come from four finite-difference schemes or from automatic
differentiation with JAX \citep{Bradbury2018}. Both routes are validated
against each other and against the reference
\href{https://github.com/brinckmann/montepython_public}{MontePython}
Euclid likelihoods \citep{Casas2024}, and a public dashboard compares
CAMB and CLASS forecasts for 25 validation cases. The survey
specifications, solver settings, source revisions and chains behind
every number quoted here are released following the FAIR principles
(findable, accessible, interoperable and reusable)
\citep{Wilkinson2016}. The change of name from \texttt{cosmicfishpie} to
\texttt{cosmicjellyfish} was motivated by the transition to JAX and the
availability of Monte Carlo methods.

\section{Statement of need}\label{statement-of-need}

Forecasts shape the analysis of a survey long before the data arrives:
they guide the choice of scale cuts and redshift binning, the modelling
of galaxy bias, intrinsic alignments and baryonic feedback, and they set
the expected sensitivity to the neutrino mass \citep{Archidiacono2025}
and to dynamical dark energy. Fisher forecasts are the standard tool
because they need tens of model evaluations instead of the \(10^5\) to
\(10^6\) of a sampler, but they rest on two assumptions: stable
numerical derivatives of the observables, and a nearly Gaussian
posterior. Neither is routinely checked, and independent Fisher
pipelines have differed by 10\% or more before their recipes were
harmonised \citep{Blanchard2020}.

\texttt{cosmicjellyfish} is built to make both checks inexpensive. The
same observable can be differentiated with four finite-difference
schemes or by automatic differentiation, and the same configuration can
be sampled with the nested sampler Nautilus \citep{Lange2023}.
Specifications for Euclid \citep{Mellier2025}, the SKA Observatory
\citep{Bacon2020} and CMB experiments \citep{Planck2018} are shipped,
transcriptions of DESI \citep{DESI2016} and Rubin/LSST
\citep{Ivezic2019} specifications are available, and further surveys are
added as YAML files. The intended users are members of survey
collaborations preparing forecast and validation studies, theorists who
need the sensitivity of a survey to a new model, and students, for whom
the package is a compact route from a power spectrum to a parameter
constraint.

\section{State of the field}\label{state-of-the-field}

Several codes produce forecasts for large-scale structure surveys.
\texttt{Augur} \citep{Rogozenski2026} is the Fisher tool of the LSST
Dark Energy Science Collaboration, built on CCL \citep{Chisari2019} and
validated against CosmoSIS \citep{Zuntz2015} and nested sampling, and
\texttt{DerivKit} \citep{Sarcevic2026} provides stable numerical
derivatives independently of any cosmology code. General inference
frameworks such as MontePython \citep{Audren2013, Brinckmann2019} and
Cobaya \citep{Torrado2021} contain forecasting likelihoods, but they are
not Fisher codes. Differentiable libraries such as \texttt{jax-cosmo}
\citep{Campagne2023}, CosmoPower-JAX \citep{Piras2023} and
\texttt{cloelib} \citep{Bonici2026} provide predictions and gradients;
\texttt{jax-cosmo} also builds Gaussian angular covariances and a Fisher
matrix by automatic differentiation, but none of them ships validated
survey specifications or a multi-probe forecasting pipeline.

Three gaps remain. First, validation: the Fisher approach for
spectroscopic surveys had to be validated against MCMC
\citep{YahiaCherif2021}, and numerical derivatives were found to
dominate the instability of LSST-like forecasts \citep{Bhandari2021}.
Second, CAMB and CLASS have been compared at the level of power spectra
\citep{Lesgourgues2011, LesgourguesTram2011}, but rarely at the level of
forecasts, where the accuracy settings for massive neutrinos matter most
\citep{Archidiacono2025}. Third, \texttt{jax-cosmo} compared automatic
and finite differences for two parameters \citep{Campagne2023}; a
controlled comparison for a full Euclid forecast with 17 cosmological
and nuisance parameters, and against an independent NumPy implementation
of the same observable, was missing.

We extended \texttt{cosmicfishpie} instead of contributing to these
frameworks for two reasons. The checks above require exchanging one
element of the forecast (theory, derivative, or Fisher against
posterior) while keeping all others fixed, so the survey specification,
observable, covariance, derivative and sampler must live in one code
base; this does not map onto a single likelihood or theory module of a
general framework. Second, the code is the reference implementation of
published Euclid forecasts
\citep{Blanchard2020, Casas2024, Archidiacono2025, Casas2026fR, Frusciante2024},
and those results had to stay reproducible. Its Fisher recipes were
validated against the MontePython likelihoods \citep{Casas2024}, which
served as a basis for current Euclid likelihood codes such as
\texttt{cloelike} \citep{cloelike2026}; \texttt{cosmicjellyfish} is
therefore an independent reference for them.

\section{Software design}\label{software-design}

A forecast is a chain of exchangeable elements: theory backend, survey
specification, observable and covariance, derivative method, and Fisher
assembly or likelihood. The design choices below serve one aim: to
change one element while keeping the others fixed, so that a difference
in the result can be attributed.

\textbf{One interface for several theory backends.} CAMB
\citep{Lewis2000}, CLASS \citep{Blas2011}, the SYREN emulators
\citep{Bartlett2024a, Bartlett2024b, Sui2025}, \texttt{jax-cosmo} and
precomputed tables are wrapped behind one \texttt{cosmo\_functions}
interface and selected with a single option. The emulators reduce a
survey forecast from minutes to seconds, at the cost of a restricted
model space (\(\Lambda\)CDM and \(w_0w_a\)CDM with massive neutrinos);
the Boltzmann solvers remain the reference, with the precision profiles
of \citet{Archidiacono2025} shipped as YAML files. New backends, such as
the differentiable DISCO-DJ \citep{Hahn2024} or CosmoPower-JAX
\citep{Piras2023}, only need to implement this interface. Beyond
\(\Lambda\)CDM, a modified-gravity module for Hu-Sawicki \(f(R)\)
gravity loads the Winther fitting formula \citep{Winther2019} and the
e-MANTIS emulator \citep{SaezCasares2024} for the nonlinear boost of the
power spectrum.

\textbf{Derivative methods replace only the Jacobian.} Finite
differences (three-point, four-point forward, polynomial fit and an
adaptive scheme) and automatic differentiation are implemented as
derivative providers that return the Jacobian of an observable; the
covariance and the Fisher assembly are shared, so the two Fisher
matrices differ only in the derivative. Automatic differentiation needs
the observables written in JAX, so the photometric and spectroscopic
observables exist in two implementations, one in NumPy and one in JAX.
The JAX version currently covers flat \(\Lambda\)CDM and \(w_0w_a\)CDM
with massless neutrinos, and the code raises an error outside this range
instead of quietly falling back to finite differences. We keep both on
purpose: each one checks the other, and their outputs are compared in
continuous integration.

\textbf{Sampling without gradients.} The likelihoods reuse the
observable and covariance code, and they are sampled with Nautilus
\citep{Lange2023}, which needs no gradients and therefore works with
CAMB and CLASS. Other samplers, such as pocoMC \citep{Karamanis2022},
and gradient-based methods such as Hamiltonian Monte Carlo with the JAX
backend, can be added through the same likelihood interface.

\textbf{Configuration as data.} Survey specifications and solver
settings are versioned YAML files, and every run writes its full
settings to a JSON file, so a forecast can be rebuilt from its output.
Every setting has a default that reproduces the Euclid
forecast-validation recipes \citep{Blanchard2020}, so a user only has to
specify the settings that should differ from the Euclid setup, as in the
example below. The former import path \texttt{cosmicfishpie} is kept as
a deprecated alias, so configurations of published forecasts still run.

\section{Example}\label{example}

The examples use version 2.1.0, archived on Zenodo (DOI for all
versions:
\href{https://doi.org/10.5281/zenodo.24593176}{10.5281/zenodo.24593176}).
The \texttt{symbolic} backend used below also needs the symbolic
emulators of the linear and nonlinear power spectrum
\citep{Bartlett2024a, Bartlett2024b}, which are not on PyPI, so the
installation is:

\begin{lstlisting}[language=bash]
pip install cosmicjellyfish==2.1.0
pip install git+https://github.com/DeaglanBartlett/symbolic_pofk.git
\end{lstlisting}

The defaults reproduce the pessimistic Euclid setup of
\citet{Blanchard2020}, so a photometric and a spectroscopic forecast,
and their combination, take three statements:

\begin{lstlisting}[language=Python]
from cosmicjellyfish.fishermatrix.cosmicfish import FisherMatrix

photo = FisherMatrix(options={"code": "symbolic", "feedback": 1},
                     observables=["WL", "GCph"], cosmoModel="LCDM").compute()
spectro = FisherMatrix(options={"code": "symbolic", "feedback": 1},
                       observables=["GCsp"], cosmoModel="LCDM").compute()
combined = photo + spectro  # nuisance parameters of both probes are kept
\end{lstlisting}

Both Fisher matrices together take about 15 s on a laptop with 16 cores.

Replacing \texttt{"symbolic"} by \texttt{"camb"} or \texttt{"class"}
repeats the forecast with a Boltzmann solver. With \texttt{"jaxcosmo"}
and a fiducial cosmology with massless neutrinos, passing
\texttt{derivative\_provider=JaxAutodiffProvider()} to
\texttt{FisherMatrix} replaces the finite differences by automatic
differentiation. The likelihood is sampled with the same survey, theory
and fiducial settings; the listing below is a minimal two-parameter
weak-lensing demonstration (the full benchmark is described in the next
section):

\begin{lstlisting}[language=Python]
from cosmicjellyfish.likelihood import NautilusSampler

config = {
    "name": "wl_demo",
    "observables": ["WL"],
    "fiducial": {"Omegam": 0.32, "Omegab": 0.05, "h": 0.67,
                 "ns": 0.96, "sigma8": 0.815584},
    "options": {"code": "symbolic", "cosmo_model": "LCDM",
                "survey_name": "Euclid", "feedback": 1},
    "priors": {"Omegam": [0.28, 0.36], "sigma8": [0.75, 0.87]},
    "sampler_settings": {"n_live": 500, "n_networks": 1, "n_batch": 50, "pool": 8},
}

if __name__ == "__main__":  # worker processes are spawned
    NautilusSampler(config).run()
\end{lstlisting}

\section{Research impact statement}\label{research-impact-statement}

\textbf{Publications.} As \texttt{cosmicfishpie}, the code has been used
in about ten publications. It produced the Euclid forecasts for the
spectroscopic and photometric probes \citep{Blanchard2020}, and was then
used in Euclid forecasts for \(f(R)\) gravity \citep{Casas2026fR}, for
scale-independent modifications of gravity \citep{Frusciante2024} and
for primordial features \citep{Ballardini2024}, in the validation of the
MontePython forecasting likelihoods \citep{Casas2024}, in the Euclid
study of the sensitivity to neutrino parameters
\citep{Archidiacono2025}, and in forecasts for radio and optical survey
synergies with the SKA Observatory \citep{Casas2023SKAO}. Earlier works
with the code and its predecessors cover modified gravity and dark
energy \citep{Casas2017, Frusciante2019, Atayde2024}, inflation and its
connection to dark energy \citep{Casas2018, Casas2019, Akrami2021}, the
anisotropic stress \citep{Sakr2025}, the validation of the Fisher
approach for spectroscopic surveys \citep{YahiaCherif2021} and the
impact of nonlinear and baryonic feedback prescriptions on weak lensing
\citep{Martinelli2021}.

\textbf{Community.} The code is developed openly on GitHub, with code
review through pull requests and continuous integration; the twelve
authors of this paper work at institutions in eleven countries.

\textbf{Validation suite.} A public dashboard
(\url{https://www.santicasas.space/cosmicjellyfish/dashboard/}) compares
CAMB and CLASS forecasts for 25 cases: the photometric and spectroscopic
Euclid probes in \(\Lambda\)CDM and \(w_0w_a\)CDM, with and without
massive neutrinos and \(N_{\mathrm{eff}}\), following \citet{Casas2024}
and \citet{Archidiacono2025}. The validation data are public and live in
a Hugging Face dataset
(\url{https://huggingface.co/datasets/santicas/cjfp-data}). With the
shipped high-precision profiles the two solvers agree to better than 1\%
on all marginalised errors of the photometric benchmark. A Euclid
photometric forecast then takes 138 s with CAMB and 186 s with CLASS,
and 30 s and 111 s with the default profiles.

\begin{figure}
\centering
\includegraphics[width=0.7\linewidth,height=\textheight,keepaspectratio]{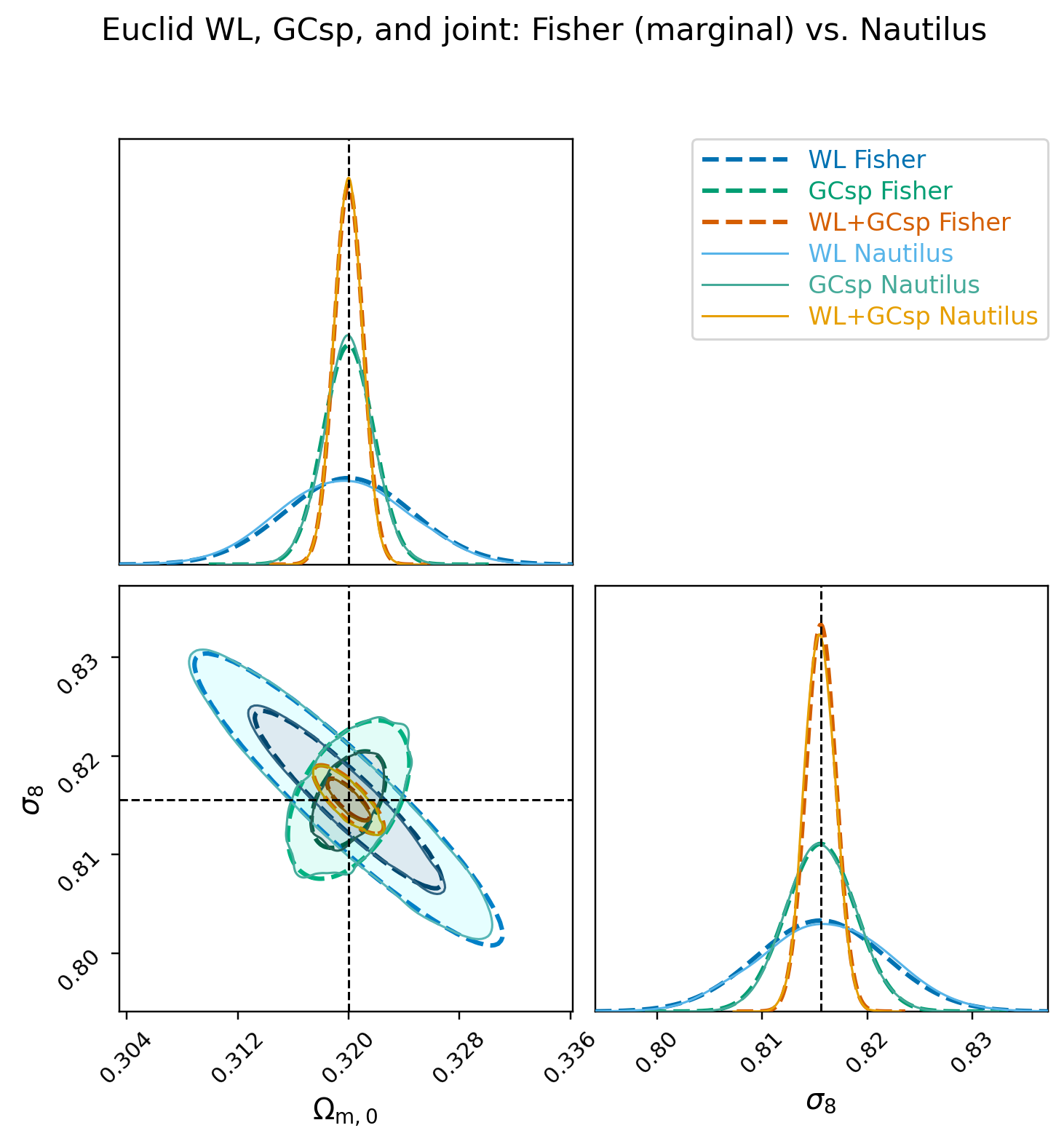}
\caption{Fisher forecasts (dashed lines) and Nautilus posteriors (thin
solid lines) for Euclid weak lensing, spectroscopic clustering and their
combination, with the symbolic backend and all nuisance parameters
marginalised in both cases. The combination assumes that WL and GCsp are
statistically independent: the log-likelihoods are summed and the Fisher
matrices added. Reproduced by
\texttt{scripts/paper-joss/reproduce\_all.sh}.\label{fig:nautilus}}
\end{figure}

\textbf{Fisher matrices against posteriors.} The Fisher matrices of the
code have been validated in several earlier works, against independent
Fisher codes \citep{Blanchard2020} and against MCMC posteriors
\citep{Casas2024, Archidiacono2025}, and the dashboard above keeps
tracking them. We repeat the comparison here with the built-in sampler.
For Euclid weak lensing (WL), spectroscopic clustering (GCsp) and their
combination, with all nuisance parameters sampled (5, 10 and 13
dimensions), the Fisher forecast and the Nautilus posterior are almost
indistinguishable (\autoref{fig:nautilus}). The marginalised errors on
\(\Omega_{\mathrm{m}}\) and \(\sigma_8\) agree to within 1.4\%, 1.2\%
and 0.9\%, which is the size of the sampling noise itself: with about
2000 to 3000 effective samples per chain, the noise on an error bar is
about 1\%. For these benchmarks, the Fisher approximation therefore
reproduces the marginalised widths of \(\Omega_{\mathrm{m}}\) and
\(\sigma_8\) to within the estimated sampling precision, and the chains,
which took 11 min, 10 min and 36 min on 64 workers, are cheap enough to
check it case by case instead of assuming it. The figure and these
numbers are produced by \texttt{scripts/paper-joss/reproduce\_all.sh}
(steps \texttt{nautilus} and \texttt{render}), which runs the three
chains and draws the comparison in about 70 min on one 64-worker node.

\begin{figure}
\centering
\includegraphics[width=0.9\linewidth,height=\textheight,keepaspectratio]{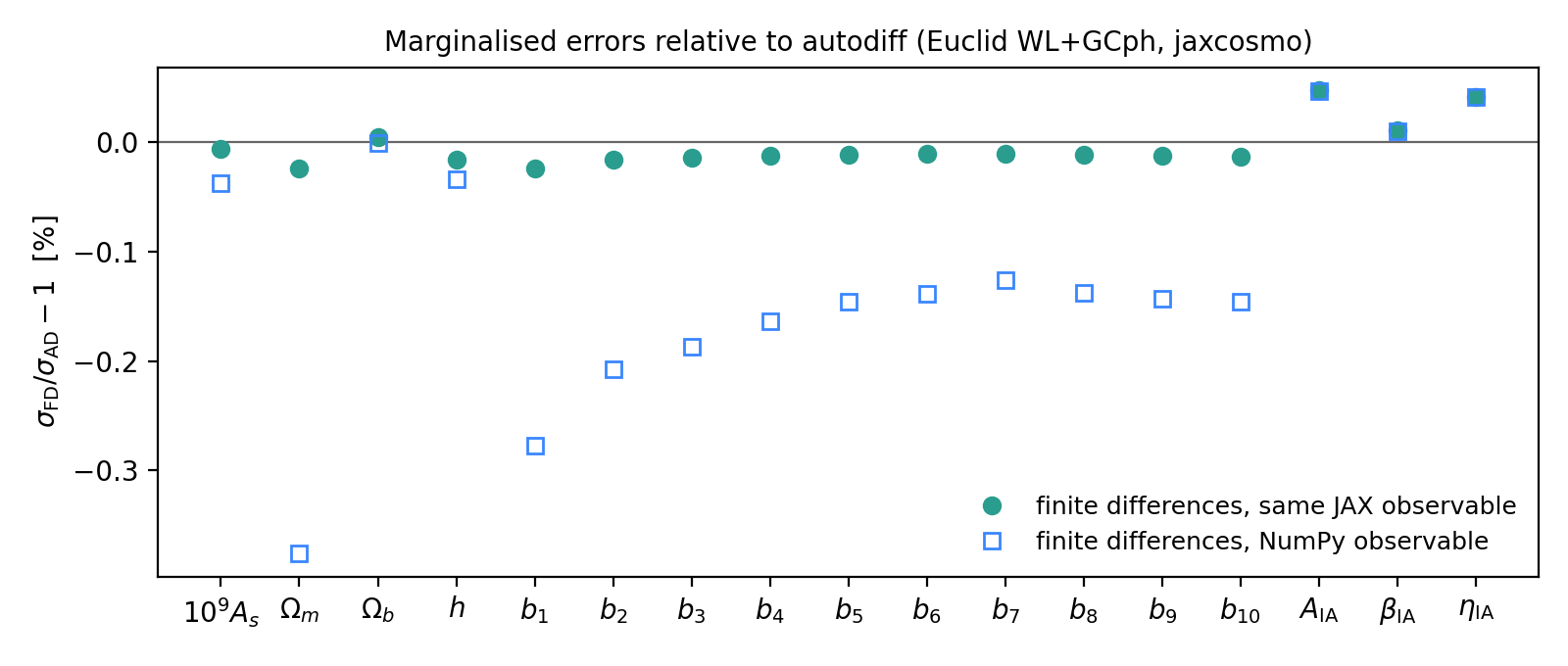}
\caption{Marginalised errors from finite differences relative to
automatic differentiation for the Euclid photometric forecast with the
\texttt{jaxcosmo} backend, for the 17 free parameters (cosmological,
galaxy bias and intrinsic alignment): same JAX observable (dots) and
standard NumPy observable (squares). Reproduced by
\texttt{scripts/paper-joss/plot\_autodiff\_ratios.py}.\label{fig:autodiff}}
\end{figure}

\textbf{Automatic differentiation.} Two Fisher matrices can differ for
two reasons: the way the derivatives are taken, and the way the
observable is computed. To separate them, we compute the Euclid
photometric Fisher matrix (WL and photometric clustering, 17 parameters)
three times: with automatic differentiation of the JAX observable, which
has no step size; with finite differences of the same JAX observable,
which isolates the derivative method; and with finite differences of the
standard NumPy implementation, which also tests the implementation of
the observable itself (\autoref{fig:autodiff}). The marginalised errors
agree to 0.05\% for the same JAX observable and to 0.4\% for the NumPy
implementation, for all 17 parameters. The JAX and NumPy implementations
are therefore equivalent for this purpose, and the finite-difference
steps are small enough not to bias the errors, so any of the three
routes can be used. These deviations are far below the 10\% or more by
which independent Fisher pipelines have differed \citep{Blanchard2020}.
The numbers come from \texttt{scripts/paper-joss/autodiff\_numbers.py}
and the figure from
\texttt{scripts/paper-joss/plot\_autodiff\_ratios.py}, both run by the
\texttt{autodiff} step of \texttt{scripts/paper-joss/reproduce\_all.sh}
in a few minutes.

\textbf{Surveys and spectroscopic settings.} The code ships
specifications for Euclid\footnote{\url{https://www.euclid-ec.org}}
\citep{Mellier2025}, DESI\footnote{\url{https://www.desi.lbl.gov}}
\citep{DESI2016}, Rubin/LSST\footnote{\url{https://rubinobservatory.org}}
\citep{Ivezic2019}, the SKA Observatory\footnote{\url{https://www.skao.int}}
\citep{Bacon2020}, MeerKLASS\footnote{\url{https://meerklass.github.io}}
\citep{Santos2017}, Planck\footnote{\url{https://www.cosmos.esa.int/web/planck}}
\citep{Planck2018}, the Simons Observatory\footnote{\url{https://simonsobservatory.org}}
\citep{Ade2019} and CMB-S4\footnote{\url{https://cmb-s4.org}}
\citep{Abazajian2016}. Euclid is provided for the photometric and
spectroscopic probes, with the pessimistic and optimistic settings of
\citet{Blanchard2020} and the data-release settings of
\citet{Archidiacono2025}; the SKA Observatory for photometric and
spectroscopic clustering and for intensity mapping, summarized in
\citet{Casas2023SKAO}; and DESI for the target samples of its
survey-validation programme \citep{DESI2024}. Spectroscopic clustering
carries more settings than the survey itself: the bias model (a linear
bias per redshift bin, parametrised directly or by its logarithm, or a
scale-dependent bias), the shot noise, the nonlinear velocity-dispersion
damping, the redshift errors and the scale cuts, with the corresponding
nuisance parameters all free in the Fisher matrix.

\begin{figure}
\centering
\includegraphics[width=1\linewidth,height=\textheight,keepaspectratio]{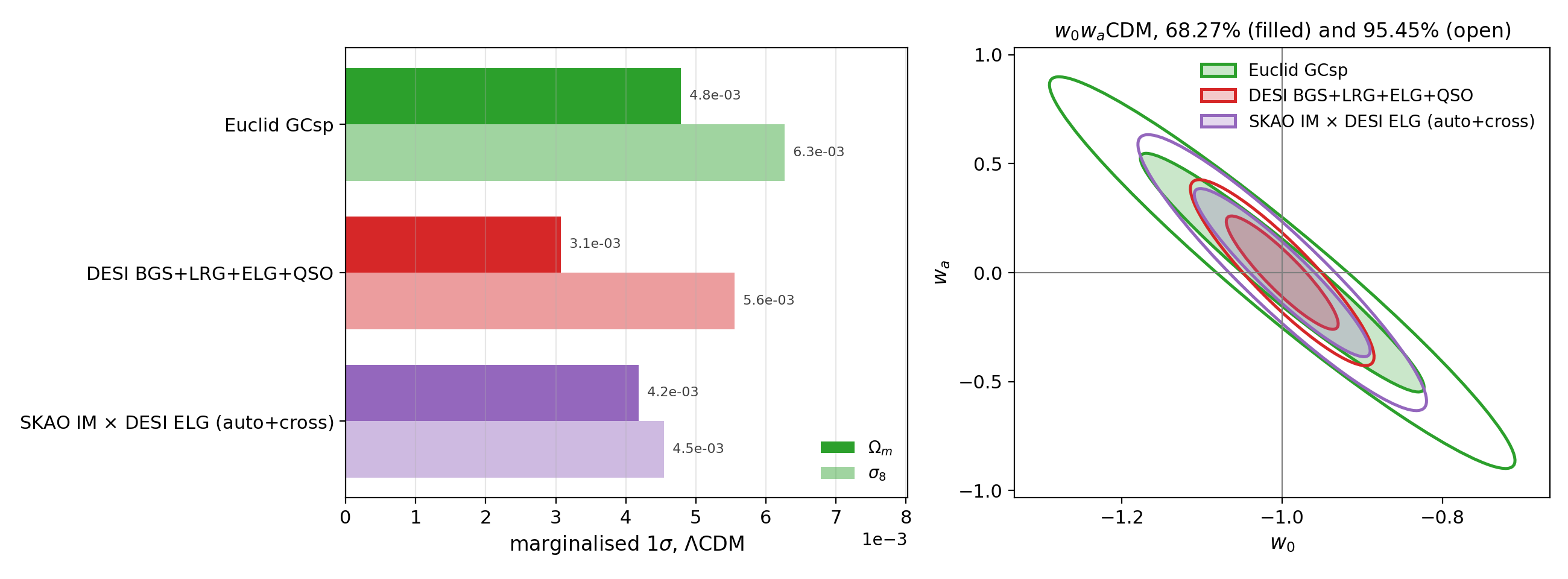}
\caption{Spectroscopic forecasts for Euclid GCsp, the four DESI samples
(independent blocks added) and SKAO intensity mapping with the DESI ELG
sample (auto- and cross-spectra), with all nuisance parameters
marginalised. Left: marginalised errors on \(\Omega_{\mathrm{m}}\) and
\(\sigma_8\) in \(\Lambda\)CDM. Right: 68.27\% and 95.45\% contours in
the \(w_0\)-\(w_a\) plane in \(w_0w_a\)CDM. Reproduced by
\texttt{paper/scripts/fig5\_spectro\_survey\_comparison.py}.\label{fig:spectro}}
\end{figure}

\autoref{fig:spectro} compares Euclid spectroscopic clustering, the four
DESI samples (BGS, LRG, ELG and QSO) and SKAO intensity mapping combined
with the DESI ELG sample, on a common \(\Lambda\)CDM fiducial and with
all nuisance parameters marginalised. The ranking reflects the
specifications as transcribed and, for DESI, the assumption that the
four samples are independent, since they cover disjoint redshift ranges.
The figure is produced by
\texttt{paper/scripts/fig5\_spectro\_survey\_comparison.py} in about 2
min.

\textbf{Modified gravity.} The code also has a modified-gravity module
for Hu-Sawicki \(f(R)\) gravity. The \(f(R)\) background and linear
growth come from a \texttt{hi\_class} \citep{Zumalacarregui2017} based
solver (\texttt{frahiclassy}, which will be made available soon), and
the nonlinear power spectrum is the \(\Lambda\)CDM one multiplied by a
boost, either the fitting formula of \citet{Winther2019} or the e-MANTIS
emulator \citep{SaezCasares2024}. The implementation follows closely the
MontePython likelihood used for the \(f(R)\) photometric analysis of
\citet{Koyama2025} (\url{https://github.com/Sefa76/photometric_fofR/}).
On the grid in multipole and redshift of that likelihood, the boost
agrees exactly (zero difference) with it, for both models and for
\(10^{-7} \le |f_{R0}| \le 10^{-4}\). The table shows the resulting
Fisher errors for the pessimistic Euclid setting (WL, photometric
clustering and their cross-correlation) in the \(f(R)\) model of
\citet{Casas2026fR}, against the Fisher forecast of that paper. The
errors on \(\Omega_\mathrm{m}\) and \(n_\mathrm{s}\) agree within 2\%
and that on \(\Omega_\mathrm{b}\) within 7\%, while those on \(h\) and
\(\sigma_8\) are 10\% to 22\% larger here; in \(\Lambda\)CDM they
coincide. Possible contributors are the sky fraction
(\(f_\mathrm{sky}=0.35\) here, 15000 deg\(^2\) \(\approx 0.364\) there)
and the scale-independent growth used here for the intrinsic alignment;
the different sky fraction alone raises the errors by only about 2\%
(\(\sqrt{0.364/0.35}\approx 1.02\)), so it explains little of the
difference. The error on \(\log_{10}|f_{R0}|\) is 0.137 with the Winther
boost and 0.128 with e-MANTIS, against 0.143 in \citet{Casas2026fR}. The
MCMC analysis of \citet{Koyama2025} gives 0.11 for both boosts, in a
different parameter basis.

\begin{longtable}[]{@{}
  >{\raggedright\arraybackslash}p{(\linewidth - 12\tabcolsep) * \real{0.3571}}
  >{\raggedleft\arraybackslash}p{(\linewidth - 12\tabcolsep) * \real{0.1071}}
  >{\raggedleft\arraybackslash}p{(\linewidth - 12\tabcolsep) * \real{0.1071}}
  >{\raggedleft\arraybackslash}p{(\linewidth - 12\tabcolsep) * \real{0.1071}}
  >{\raggedleft\arraybackslash}p{(\linewidth - 12\tabcolsep) * \real{0.1071}}
  >{\raggedleft\arraybackslash}p{(\linewidth - 12\tabcolsep) * \real{0.1071}}
  >{\raggedleft\arraybackslash}p{(\linewidth - 12\tabcolsep) * \real{0.1071}}@{}}
\caption{Fisher forecast for the \(f(R)\) model of \citet{Casas2026fR},
Euclid pessimistic setting: marginalised relative errors (\%) on the
\(\Lambda\)CDM parameters and absolute error on
\(\log_{10}|f_{R0}|\).}\tabularnewline
\toprule\noalign{}
\begin{minipage}[b]{\linewidth}\raggedright
Error
\end{minipage} & \begin{minipage}[b]{\linewidth}\raggedleft
\(\Omega_\mathrm{m}\) (\%)
\end{minipage} & \begin{minipage}[b]{\linewidth}\raggedleft
\(\Omega_\mathrm{b}\) (\%)
\end{minipage} & \begin{minipage}[b]{\linewidth}\raggedleft
\(h\) (\%)
\end{minipage} & \begin{minipage}[b]{\linewidth}\raggedleft
\(n_\mathrm{s}\) (\%)
\end{minipage} & \begin{minipage}[b]{\linewidth}\raggedleft
\(\sigma_8\) (\%)
\end{minipage} & \begin{minipage}[b]{\linewidth}\raggedleft
\(\sigma(\log_{10}\lvert f_{R0}\rvert)\)
\end{minipage} \\
\midrule\noalign{}
\endfirsthead
\toprule\noalign{}
\begin{minipage}[b]{\linewidth}\raggedright
Error
\end{minipage} & \begin{minipage}[b]{\linewidth}\raggedleft
\(\Omega_\mathrm{m}\) (\%)
\end{minipage} & \begin{minipage}[b]{\linewidth}\raggedleft
\(\Omega_\mathrm{b}\) (\%)
\end{minipage} & \begin{minipage}[b]{\linewidth}\raggedleft
\(h\) (\%)
\end{minipage} & \begin{minipage}[b]{\linewidth}\raggedleft
\(n_\mathrm{s}\) (\%)
\end{minipage} & \begin{minipage}[b]{\linewidth}\raggedleft
\(\sigma_8\) (\%)
\end{minipage} & \begin{minipage}[b]{\linewidth}\raggedleft
\(\sigma(\log_{10}\lvert f_{R0}\rvert)\)
\end{minipage} \\
\midrule\noalign{}
\endhead
\bottomrule\noalign{}
\endlastfoot
Table 2 of \citet{Casas2026fR} & 0.80 & 5.0 & 3.3 & 1.7 & 0.40 &
0.143 \\
\texttt{cosmicjellyfish}, Winther boost & 0.81 & 5.34 & 3.64 & 1.74 &
0.48 & 0.137 \\
\texttt{cosmicjellyfish}, e-MANTIS boost & 0.82 & 5.37 & 3.72 & 1.76 &
0.49 & 0.128 \\
\end{longtable}

The configurations, source revisions and chains behind every number in
this section are public
(\url{https://huggingface.co/datasets/santicas/cjfp-data}). The \(f(R)\)
table, however, cannot yet be rerun from public inputs alone, since it
needs the \texttt{frahiclassy} solver, which is not public yet.

\section{AI usage disclosure}\label{ai-usage-disclosure}

Over the last year, several large language models were used as coding
assistants for the code, tests and documentation: OpenAI GPT models,
from GPT-5 to, most recently, GPT-6.1 Astra, Sol and Luna; Anthropic
Claude models, mostly Claude Opus 5.5 and Claude Sonnet 5.5 and earlier
versions; and many open-weight models served kindly by Blablador, the
LLM service of the Jülich Supercomputing Centre for Helmholtz AI
(\url{https://sdlaml.pages.jsc.fz-juelich.de/ai/guides/blablador_api_access/}),
mostly DeepSeek-V4.1-Flash, but also Mistral, Kimi and Muse models. The
repository ships an \texttt{AGENTS.md} guide for coding agents. All
AI-assisted changes to the code, tests and documentation went through
pull requests reviewed by the authors, and were checked by the unit
tests, the continuous integration and the validation suite described
above.

\begin{figure}
\centering
\includegraphics[width=1\linewidth,height=\textheight,keepaspectratio]{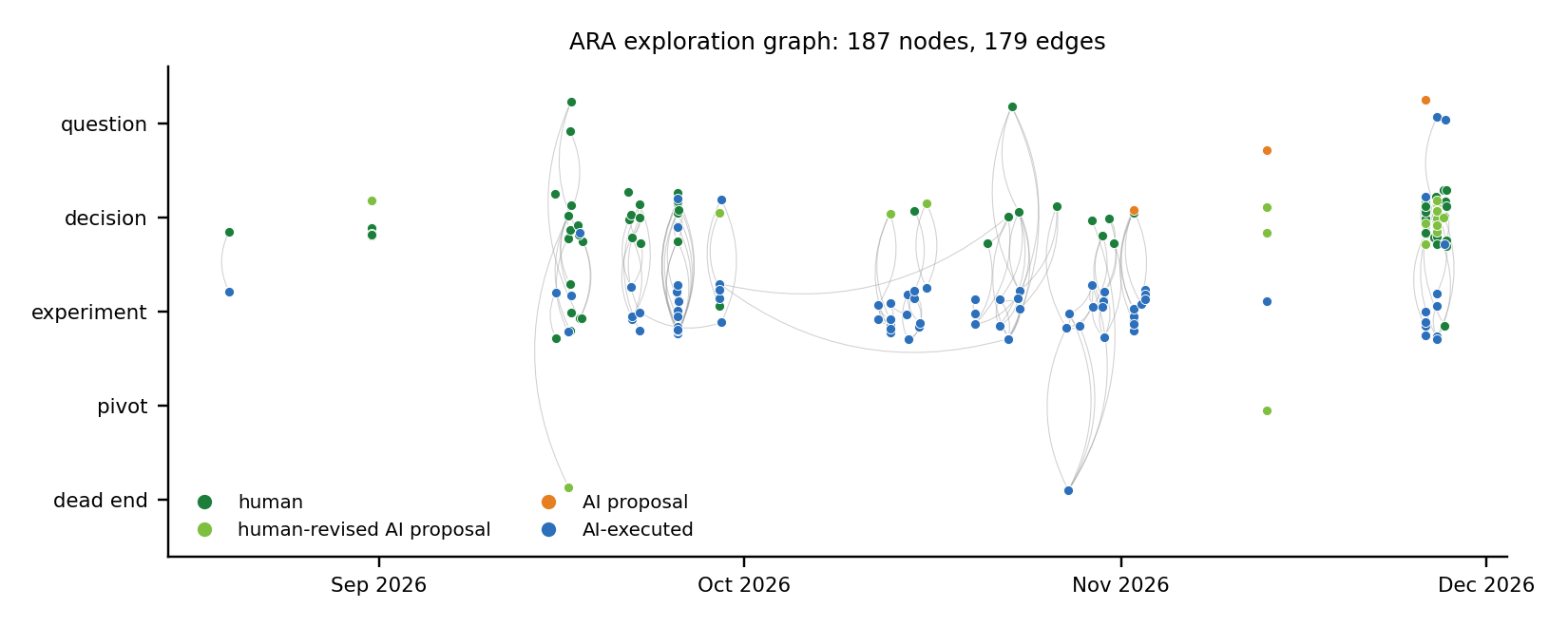}
\caption{Exploration graph of the ARA trace of \texttt{cosmicjellyfish}.
Each point is a node, placed by date and node type and coloured by
provenance; grey lines connect a node to the nodes it depends on.
Produced with
\texttt{paper/scripts/fig6\_ara\_trace.py}.\label{fig:ara}}
\end{figure}

The development decisions were recorded with the Agent-Native Research
Artifact (ARA) protocol \citep{Liu2026ARA}
(\url{https://github.com/ARA-Labs/Agent-Native-Research-Artifact}). At
the end of each working session with a coding agent, the questions,
decisions, experiments, pivots and dead ends are written to an
exploration graph, each tagged with its provenance: proposed and
executed by a human, proposed by the agent, or proposed by the agent and
revised by a human. The trace is public on the \texttt{dev} branch of
the repository, and \autoref{fig:ara} shows it for the last months of
development.

For this manuscript, the authors wrote the first draft of the Summary,
Statement of need, State of the field, Software design and the
validation text. Claude (Anthropic; mostly Claude Opus 5.5) was used to
correct the language, to restructure the text to the JOSS guidelines, to
propose and format references, to draft the Research impact statement
from the repository documentation, to test the code listings against the
repository. The authors reviewed, edited and validated all of this
content.

\section{Acknowledgements}\label{acknowledgements}

We thank Matteo Martinelli for initiating this project, the ideas, the
motivation and for bringing researchers together.

We thank Valeria Pettorino, Julien Lesgourgues, Nils Schöneberg and
Maria Archidiacono for scientific ideas, validation and discussions.

DS acknowledges support from the Fondecyt Regular project N.1251339
(ANID). ML is supported by grant PIP 11220200100729CO CONICET and grant
20020170100129BA UBACYT.

We acknowledge the use of computing resources at the JURECA cluster of
the Forschungszentrum Jülich (FZJ) under the project name paj2526 and
punch\_astro. We acknowledge the Blablador service of Helmholtz AI at
the Jülich Supercomputing Centre for access to open-weight large
language models.

\section{Author contributions}\label{author-contributions}

Author contributions follow the CRediT (Contributor Roles Taxonomy)
\citep{Brand2015}. For the authors who had not specified their roles, we
derived a first proposal from the all-contributors record of the
repository (\texttt{code} as Software, \texttt{ideas} as
Conceptualization, \texttt{design} as Methodology, \texttt{research} as
Investigation, \texttt{test}, \texttt{bug}, \texttt{review} and
\texttt{userTesting} as Validation, \texttt{data} as Data curation,
\texttt{projectManagement} and \texttt{maintenance} as Project
administration, \texttt{mentoring} as Supervision, \texttt{doc},
\texttt{example} and \texttt{tutorial} as Software, and \texttt{content}
as Writing -- original draft).

\begin{itemize}
\tightlist
\item
  \textbf{S. Casas}: Conceptualization, Data curation, Investigation,
  Methodology, Project administration, Software, Supervision,
  Validation, Visualization, Writing -- original draft, Writing --
  review \& editing.
\item
  \textbf{S. Pamuk}: Conceptualization, Investigation, Methodology,
  Software, Validation, Writing -- original draft.
\item
  \textbf{V. M. Sabarish}: Conceptualization, Investigation, Software.
\item
  \textbf{Z. Sakr}: Conceptualization, Investigation, Methodology,
  Software, Validation.
\item
  \textbf{D. Sapone}: Conceptualization, Software, Supervision.
\item
  \textbf{M. Leizerovich}: Conceptualization, Software, Validation.
\item
  \textbf{N. Frusciante}: Methodology, Software, Data curation.
\item
  \textbf{F. Pace}: Methodology, Software, Data curation.
\item
  \textbf{K. Koyama}: Conceptualization, Methodology, Supervision,
  Validation.
\item
  \textbf{E. Bellini}: Conceptualization, Data curation, Software.
\item
  \textbf{S. Camera}: Conceptualization, Methodology.
\item
  \textbf{M. Raveri}: Conceptualization, Software.
\end{itemize}

\bibliographystyle{mnras}
\bibliography{paper}

\end{document}